\documentclass{aa}
\usepackage{psfig,longtable,lscape}
\newcommand{\et}{et al.}
\newcommand{\ct}{cts s$^{-1}$}
\newcommand{\perr}{r$_{90}$}

\newcommand{\ext}{r$_{\rm ext}$}

\newcommand{\extl}{ML$_{\rm ext}$}
\newcommand{\exil}{ML$_{\rm exi}$}

\begin{document}

\thesaurus{
           04.03.1; 
           11.13.1; 
           11.19.5; 
           13.25.2; 
           13.25.5 
          }

\title{ROSAT HRI catalogue of X-ray sources in the LMC
region\thanks{Table 4 is only available in electronic form
at the CDS via anonymous ftp to cdsarc.u-strasbg.fr (130.79.128.5)
or via http://cdsweb.u-strasbg.fr/Abstract.html}
}

\author{Manami Sasaki, Frank Haberl, and Wolfgang Pietsch}
\authorrunning{Sasaki et al.}

\offprints{M.\ Sasaki (manami@mpe.mpg.de)}

\institute{Max--Planck--Institut f\"ur extraterrestrische Physik,
 Giessenbachstra{\ss}e, 85748 Garching, Germany}

\date{Received date: 01 December 1999; accepted date: 18 January 2000}

\maketitle

\begin{abstract}
All 543 pointed observations of the ROSAT High Resolution Imager (HRI)
with exposure times higher than 50 sec
and performed between 1990 and 1998 in a field of 10\degr\ x
10\degr\ covering the Large Magellanic Cloud (LMC) were analyzed.
A catalogue was produced containing 397 X-ray sources with
their properties measured by the HRI. The list was cross-correlated
with the ROSAT Position Sensitive Propotional 
Counter (PSPC) source catalogue presented by Haberl \& Pietsch (1999)
in order to obtain the hardness ratios for the X-ray sources detected
by both instruments. 138 HRI sources are contained in the PSPC
catalogue, 259 sources are new detections. 
The spatial resolution of the HRI was
higher than that of the PSPC and the source position could be
determined with errors  mostly smaller than 15\arcsec\ which are
dominated by systematic attitude errors.  
After cross-correlating the source catalogue with the SIMBAD data base
and the TYCHO catalogue 94 HRI sources were identified with 
known objects based on their positional coincidence and X-ray
properties. Whenever more accurate coordinates were given in
catalogues or literature for identified sources, the X-ray coordinates
were corrected and the systematic error of the X-ray position was
reduced. For other sources observed simultaneously with an identified
source the coordinates were improved as well. 
In total the X-ray position of 254 sources could be newly determined.
The catalogue contains 39 foreground stars, 24 supernova remnants
(SNRs), five supersoft sources (SSSs), nine X-ray binaries (XBs), and
nine AGN well known from literature. Another eight sources
were identified with known candidates for these source classes. 
Additional 21 HRI sources are suggested in the present work as
candidates for SNR, X-ray binary in the LMC, or background AGN
because of their extent, hardness ratios, X-ray to optical flux ratio,
or flux variability.   

\end{abstract}

\section{Introduction}

The Magellanic Clouds (MCs) as the nearest galaxies to the Milky Way
allow us to resolve their stellar content in various wavelength
bands.  
X-ray observations combined with optical and radio data can be used 
to investigate the physical properties of individual X-ray sources as 
well as the statistical properties of different source classes in a 
galaxy as a whole. The quantitative and positional distribution of 
X-ray sources in the MCs will help us to understand the unresolved
X-ray emission from more distant galaxies.

After the first observation of X-ray emission from the MCs in 1968 
(Mark et al.\ 1969) four permanent (LMC X-1, X-2, X-3, and X-4, Leong
et al.\ 1971; Giacconi et al.\ 1972) and
few transient X-ray sources were found in the LMC in several satellite
missions (UHURU, SAS-3, Copernicus, Ariel-V, HEAO-1). 
An extensive pointed survey of the LMC was performed
by the Einstein Observatory between 1979 and 1981. The two detectors
on board this satellite, the Imaging Proportional Counter and the High
Resolution Imager, were sensitive enough to detect 
X-ray binaries, SSSs, and SNRs at the distance of the LMC (55kpc). 
Long et al.\ (1981)
published a list of 97 discrete X-ray sources in the direction of the
LMC and the same data was re-analyzed by Wang et al.\ (1991) finally
giving a list of 105 sources. 54 discrete X-ray sources were
identified with objects in the LMC, most of the remaining sources were
associated with foreground stars and background AGN. In EXOSAT
observations few additional X-ray sources were found (Jones et al.\
1985; Pakull et al.\ 1985; Pietsch et al.\ 1989). 

The next thorough survey of the LMC was made by ROSAT in the energy
range of 0.1 -- 2.4 keV (Tr\"umper 1982). From 1990 to 1998 ROSAT
performed more than 700  
pointed observations in a 10 by 10 degree field centered on the
LMC. Haberl \& Pietsch (1999b, hereafter HP99b) analyzed 212 PSPC
observations and created a catalogue of 758 X-ray sources. 

In this work results of the analysis of the ROSAT HRI data of the LMC
are presented. A description of the HRI detector can be found in David
\et\ (1996). A source catalogue was obtained in a
similar way as in  
HP99b and many sources were identified by cross-correlating
the source list with other existing catalogues. With the help of 
known properties of different source classes we looked for new
candidates for SNRs, stars, and hard X-ray sources which
mainly consist of X-ray binaries and absorbed background AGN.

\section{ROSAT HRI data}

\subsection{Data analysis}\label{data}

\begin{figure}[t]
\centerline{\psfig{figure=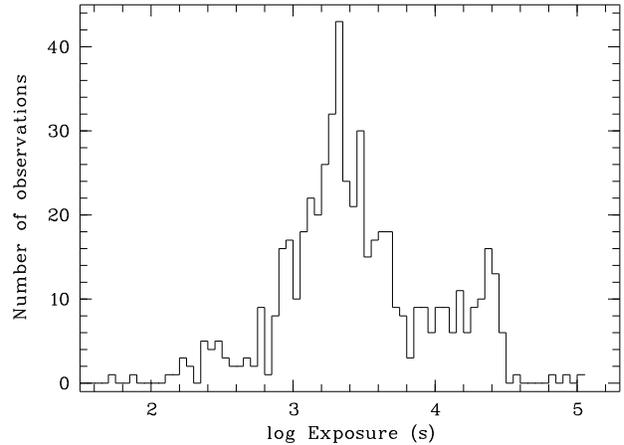,angle=270,width=9cm}}
\caption[]{\label{exphisto} Histogram of HRI pointing exposure times.}
\end{figure}

The LMC was observed by the ROSAT HRI in more than 500
pointings during the operational phase of ROSAT between 1990 and
1998. 543 observations with exposure
times of 50 to 110000 s (Fig.\,\ref{exphisto}) in a field of 10\degr\ x
10\degr\ around RA = 05$^h$ 25$^m$ 00$^s$, Dec = --67\degr\ 43\arcmin\
20\arcsec\ (J2000.0) were used for the analysis. The analysis was 
carried out using three detection methods available in EXSAS (Zimmermann et
al.\ 1994). For each pointing X-ray sources were searched using the sliding
window methods with local background and with a spline
fitted background map. The resulting detection lists were merged and  
a maximum likelihood algorithm was performed on this list.
Sources were accepted if their likelihood of existence was larger than
10.0, i.e.\ the existence probability was higher than P = 1 --
exp(--\exil) = 1 -- 4.5 $\cdot$ 10$^{-5}$, and their telescope
off-axis angle smaller than 15\arcmin\ during the observation.   

For point and point like sources the source extent was determined
by the maximum likelihood technique fitting the source intensity
distribution with a Gaussian profile. The count rates resulting from this
calculation are correct only for sources with small extent and a
brightness profile peaking in the center. For extended sources like
SNRs with ringlike structure the net count rates were determined
interactively by integrating the counts within a circle around the
source. For the background the counts were averaged in a ring around
the source distant enough not to be influenced by the source emission. 

In order to increase the sensitivity HRI observations
with pointing directions within a radius of 1\arcmin\ were
merged after adjusting their position. 
This was possible for 56 different regions in the LMC.
Source detection was also performed on these data and additional
faint sources were found which were not detectable in single
pointings.

The final source lists obtained for each pointing and co-added
observations were merged to one list and multiple detections of a source
were reduced to one detection for each source. For this purpose the
detection with the smallest positional error was chosen. After
screening manually in order to delete spurious detections like knots
in extended emission, the catalogue finally contains 397 distinct sources. 

\subsection{Positional corrections and error} \label{poscorr}

\begin{table*}[t]
\caption[]{HRI sources with significant flux variability}
\begin{tabular}{rrrrrrrl}
\hline\noalign{\smallskip}
\multicolumn{1}{c}{1} & \multicolumn{1}{c}{2} & \multicolumn{1}{c}{3} & \multicolumn{1}{c}{4} &
\multicolumn{1}{c}{5} & \multicolumn{1}{c}{6} & \multicolumn{1}{c}{7} & ~~8 \\
\hline\noalign{\smallskip}  
\multicolumn{1}{c}{No} & \multicolumn{1}{c}{Rate} &
\multicolumn{1}{c}{Rate} & 
\multicolumn{1}{c}{$\frac{\rm F_{max}}{\rm F_{min}}$} &
\multicolumn{1}{c}{Red.\ $\chi^2$} & \multicolumn{1}{c}{DOF} & 
\multicolumn{1}{c}{No} & Remarks \\
 & \multicolumn{1}{c}{HRI} & \multicolumn{1}{c}{PSPC} & &
 & & \multicolumn{1}{c}{PSPC} & \\
 & \multicolumn{1}{c}{[\ct]} & \multicolumn{1}{c}{[\ct]} & &
 & & & \\
\noalign{\smallskip}\hline\noalign{\smallskip}
 19 & *4.6e-1 & *2.3e-3 &  598.1 &376.1 & 1 & 331 & HMXB RX J0502.9-6626 (CAL E)\\ 
 20 & 5.3e-2 & 7.7e-2 & 4.5 &163.5 & 2 & 380 & AGN RX J0503.1-6634, z=0.064 [SCF94]\\ 
 23 & 2.4e-2 & 2.1e-2 & 3.4 & 61.9 & 1 & 715 &  \\ 
 49 & *4.9e-3 & *7.5e-3 & 4.3 &9.4 & 4 & 559 & $<$XB$>$ or $<$AGN$>$\\ 
 65 & 2.2e-3 & 1.9 & 25611.5 &663.9 & 45 & 1030 & SSS RX J0513.9-6951  \\ 
103 & 4.1e-3 & 2.8e-3 &10.5 & 10.2 & 6  && foreground star HD 35862  \\ 
124 & 1.8e-2 & 1.0e-1 & 4.9 & 19.6 & 20 & 1094 & AGN RX J0524.0-7011, z=0.151 [SCF94]\\ 
155 & *8.3e-3 & *4.2e-3 &  132.1 & 18.7 & 30 && Nova LMC 1995 [OG99] \\ 
167 & 1.4e-3 & 1.2e-1 &  256.1 & 51.6 & 26 & 1039 & SSS RX J0527.8-6954  \\ 
180 & *1.9 & *7.5 &11.2 &691.6 & 74 & 122 & foreground star K1III\& HD 36705 (AB Dor)\\ 
193 & 3.4e-2 & 1.7e-1 & 2.6 &7.0 & 8 & 749 & foreground star G5 HD 269620 [CSM97]\\ 
202 & 2.0e-3 & 8.5e-2 & 1022.5 & 11.9 & 20 & 204 & HMXB Be/X 
RXJ0529.8-6556 [HDP97]\\ 
218 & 8.4e-3 & 3.7e-1 &  344.9 & 99.9 & 10 & 252 & HMXB Be/X 
EXO053109-6609 [HDP95a], [DHP96]\\ 
233 & 6.0e-3 & 2.2e-2 &66.1 &7.9 & 13 & 184 & HMXB RX J0532.5-6551 (Sk -65 66) [HPD95b]\\ 
239 & *8.9e-2 & *4.9 &  367.1 & 10247.6 & 25 & 316 & HMXB LMC X-4, HD 269743 O8III  \\ 
293 & 4.4e-2 & 1.6e-1 & 2.1 &9.4 & 18 & 902 & foreground star dMe CAL 69 [CSM97]  \\ 
300 & 5.4e-3 & &18.9 &414.6 & 2 && $<$stellar$>$, source not resolved by the PSPC\\ 
306 & *6.0 & *23.4 & 2.5 &  1838.8 & 22 & 41 & HMXB LMC X-3 \\ 
311 & 3.5 & 13.5 & 1.6 &576.1 & 29 & 1001 & HMXB LMC X-1, O8III  \\ 
313 & *6.3e-3 & *8.3e-3 & 3.7 &9.3 & 3 & 668 & $<$stellar$>$\\ 
348 & 4.3e-2 & &  284.0 &775.9 & 6 & 654 & SSS CAL 83 [SCF94], one PSPC point., source near rim  \\ 
349 & 3.0e-2 & 1.0e-1 &19.5 &7.2 & 5 & 61 & foreground star? [HP99a]  \\ 
352 & 1.3e-3 & 1.2e-2 & 3.1 & 10.0 & 2 & 1225 & HMXB RX J0544.1-7100 [HP99b]\\ 
363 & 6.1e-2 & 1.3e-1 & 1.5 & 36.8 & 3 & 1240 & SSS CAL 87 \\ 
364 & 1.2e-2 & &30.5 &158.4 & 1 & 747 & $<$XB$>$ or $<$AGN$>$, one PSPC pointing, source near rib \\ 
375 & 3.3e-3 & 3.3e-2 & 4.5 &6.1 & 4 & 1127 & foreground star F3/F5IV/V HD 39756  \\ 
\noalign{\smallskip}
\hline
\end{tabular}
\label{variable}

\vspace{2mm}
Notes to columns No 2 and 3:
For point and point like sources count rates are the mean of output
values from maximum likelihood algorithm for single pointings. For
extended sources and bright sources with apparent extent (see text)
the average of integrated count rates in single pointings was taken (*
in front of the number).  

\vspace{1mm}
Notes to column No 6:
Degrees of freedom.

\vspace{1mm}
Notes to column No 7:
Source number from HP99b.

\vspace{1mm}
Notes to column No 8:
Sources classified in this work are put in $<$ $>$.
Abbreviations for references in square brackets are given in
literature list.
\end{table*}

ROSAT observations suffer from a systematic
positional uncertainty of about 7\arcsec\ (K\"urster 1993). 
For minimizing this systematic error the coordinates of
identified objects were compared to high accuracy positions available
in the TYCHO catalogue obtained from the ESA Hipparcos space
astrometry satellite (Hoeg \et\ 1997) or in the literature. 
First the X-ray position was corrected to TYCHO coordinates. For
sources without any TYCHO counterpart, but identified on the ESO
Digitized Sky Survey 
(DSS) frame with other stars on this frame which were listed in the
TYCHO catalogue, more accurate coordinates were calculated for HRI 
sources by determining the offset between the TYCHO and DSS positions
and between the HRI and DSS position. Other sources could be
identified with objects in the SIMBAD data base operated at the Centre
de Donn\'ees astronomiques de Strasbourg or in the literature and
their positions were corrected after checking their positions on DSS
frames. 
Correction of coordinates for one source implied improved coordinates
for all detections of this source in different pointings
and for other sources in same pointings.
Those secondary corrections again allowed correction of further
pointings if the sources were detected several times.
Finally for 254 out of 397 sources improved coordinates were 
determined. 

In cases where positional correction was possible the remaining
systematic error consists of the error in former optical measurements
and the statistical error of the identified source. 
For not corrected sources the systematic error was set to 7\arcsec. 
The positional error was finally computed as a composite of the
statistical uncertainty with 90 \% confidence and the systematic error.
It is used throughout the paper for the error circle. 
After the source detection procedure the mean positional error was 8\farcs3. 
The coordinate correction reduced the mean positional error of all
sources to 6\farcs4. For position corrected sources the mean
positional error is 5\farcs1. 

\subsection{Correlation with existing catalogues} \label{correlate}

\begin{figure*}[t]
\begin{center}
\begin{tabular}{lll}
\hspace{-1mm}\mbox{\psfig{figure=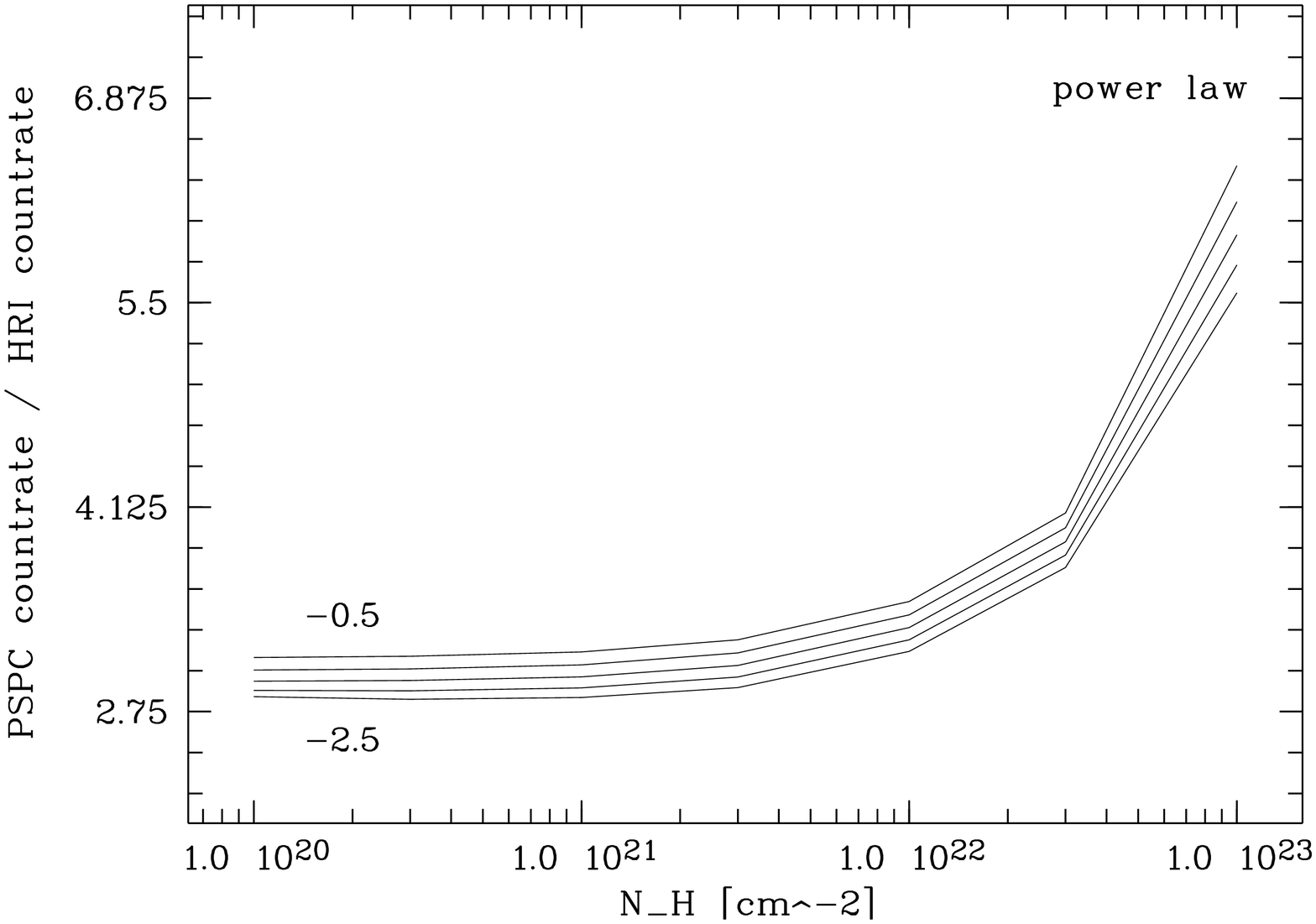,angle=270,width=6.1cm}}& 
\hspace{-6mm}\mbox{\psfig{figure=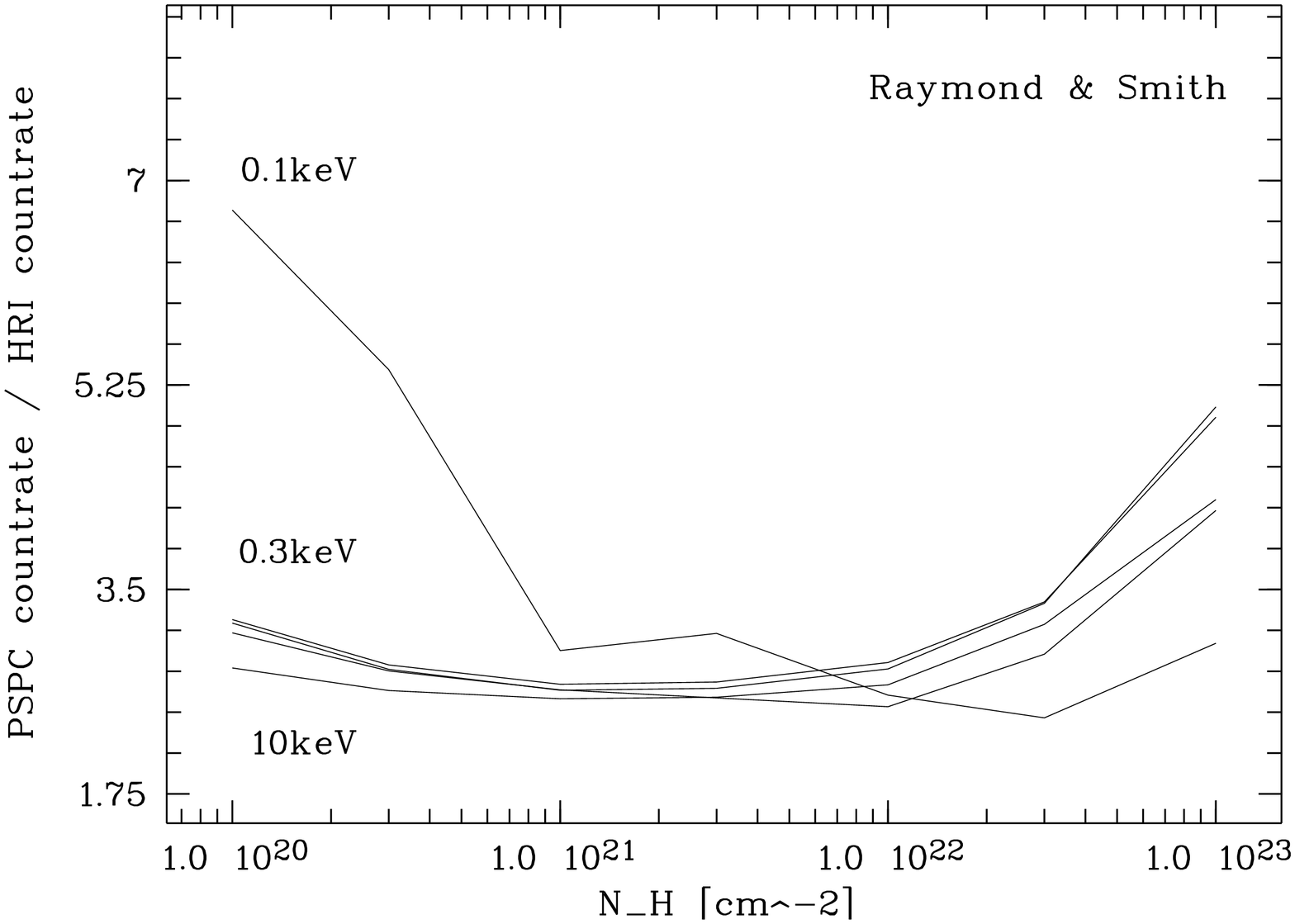,angle=270,width=6.1cm}}&
\hspace{-7mm}\mbox{\psfig{figure=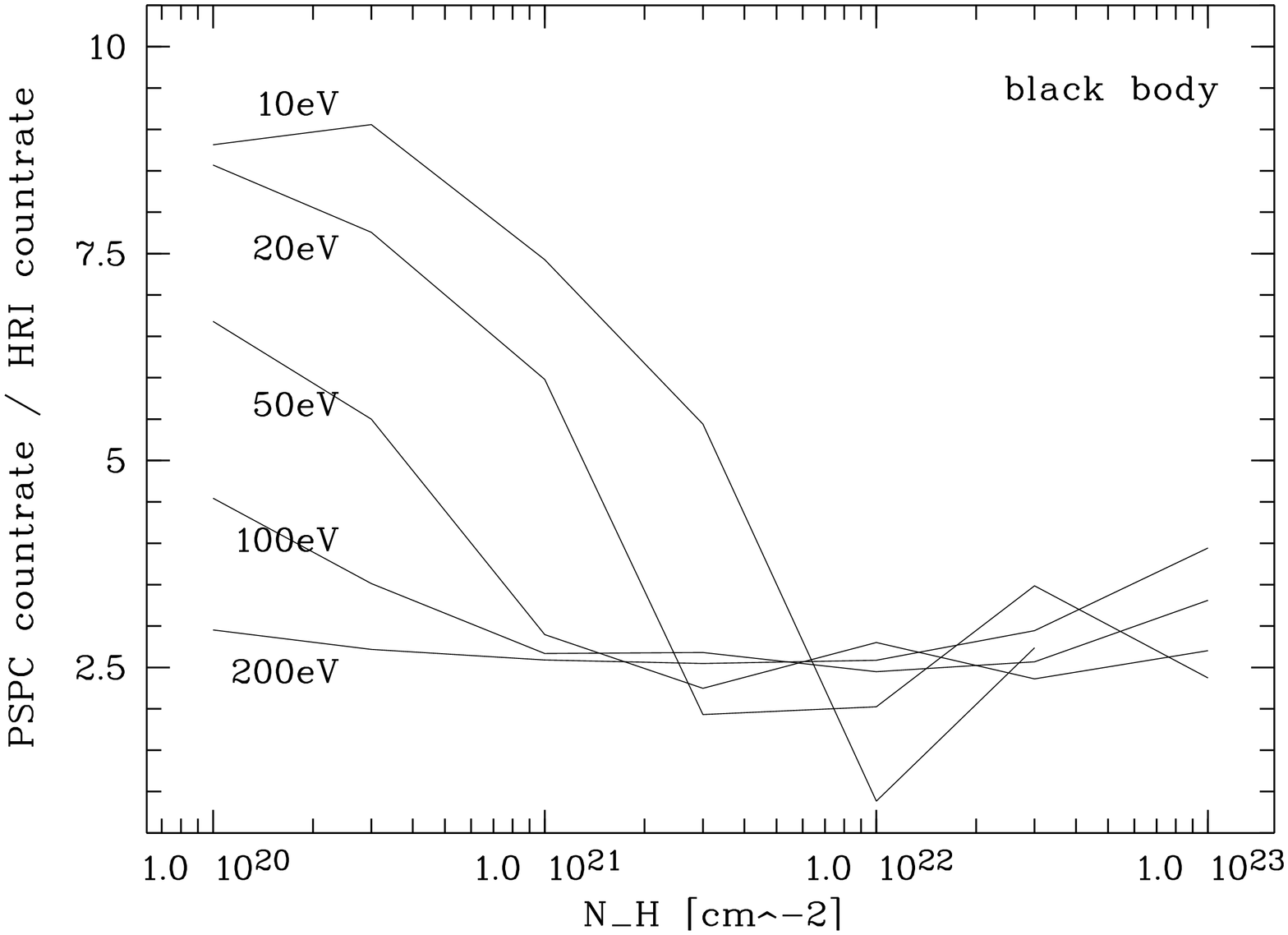,angle=270,width=6.1cm}}\\
\hspace{7mm} a.\ power law spectra with photon&
\hspace{2mm} b.\ Raymond \& Smith spectra with&
\hspace{1mm} c.\ black body spectra with\\
\hspace{7mm} index: --2.5, --2.0, --1.5, --1.0, --0.5 &
\hspace{2mm} T = 0.1, 0.3, 1.0, 3.0, 10.0 keV &
\hspace{1mm} T = 10.0 -- 200.0 eV \\
\end{tabular}
\end{center}
\vspace{-3mm}
\caption[]{\label{hri2pspc} PSPC/HRI conversion factor as 
function of N$_{\rm H}$ for power law, Raymond \& Smith, and black
body spectra.} 
\end{figure*}

The catalogue was cross-correlated with the SIMBAD data
base and the TYCHO catalogue in order to identify HRI
sources. The HRI catalogue contains samples of known SSSs,
X-ray binaries, SNRs, Galactic foreground stars, and background AGN. 
The catalogue was also cross-correlated with the source
list from the pointed PSPC observations (HP99b). 138 HRI sources are
identical with sources which were detected in PSPC data and thus for
most of them the hardness ratios (HR1, HR2) are known. Since the HRI
had no spectral resolution no information on the X-ray spectrum could
be obtained for HRI sources which are completely new detections. 
A total of 94 HRI sources were identified with known
objects like SSSs, X-ray binaries, SNRs, stars, and background AGN. 

With the help of their X-ray properties like extent,
extent likelihood, PSPC hardness ratios, X-ray to optical flux ratio
(see Sec.\,\ref{newclass}), and X-ray variability 14 previously unknown
HRI sources and 7 sources also listed in the PSPC catalogue were newly
classified.  

The whole source catalogue from HRI observations with the corrected
coordinates, final positional error, existence likelihood, HRI
count rate, extent, extent likelihood, PSPC count rate and the
corresponding PSPC source number with hardness ratios (HP99b) is given
in Table \ref{wholecat}.  
For each HRI and PSPC count rate the results for the pointing with the
smallest positional error, determined by the maximum likelihood
algorithm, were selected. 
Therefore HRI count rates in the table are representative for one single
observation for each source. For extended SNRs the given count rate may
correspond to a knot within the source.
PSPC count rates are taken from the PSPC catalogue (HP99b) if
available. For HRI sources without PSPC detection we derived
2$\sigma$ upper limit from the pointing with the highest exposure time. 
If the source was too close to the rim or the window support structure
of the PSPC detector, no count rate is given in Table \ref{wholecat}.
Neither was it possible to determine PSPC count rates or upper limits
for sources located in regions with diffuse emission. 

\subsection{Flux variability}

About 80 \% of HRI sources were observed more than once and allow time
variability studies. For point and point like sources longterm
lightcurves were produced with observation-average count rates or
upper limits determined by the maximum likelihood algorithm, whereas
for extended sources integrated count rates within a circle were used
(see Sec.\,\ref{data}). 
For some very bright sources the count rates were integrated in the
same way, because an apparent extent resulted from the maximum
likelihood algorithm. An apparent extent is computed if the high photon
statistics of the bright sources cause a significant deviation from
the assumed model for the point spread function.

A $\chi^{2}$-test for a constant count rate was performed and the
factor between the maximum and minimum flux was computed for each
lightcurve. Together with the reduced $\chi^{2}$ this flux factor was
used to characterize variability on long time scales of days to
years (see also HP99a). For SNRs we expect constant integrated flux,
however the flux factor was in the range of 1.0 to 1.8. This may be
caused by different off-axis angles and/or different extraction of the
extended source. Therefore variations below a
factor of 2.0 should be handled with care as they might indicate no
real variability but false integration of the source flux because of
the extent or existence of a nearby bright source.

In order to obtain a complete lightcurve of the ROSAT observations,
also PSPC count rates and upper limits were calculated for the HRI
sources. In Figures \ref{hri2pspc} a -- c the
PSPC to HRI count rate conversion factor is plotted over N$_{\rm H}$ =
10$^{20}$ -- 10$^{23}$ cm$^{-2}$ for three different spectral models. 
SSSs with a soft black body spectrum can be modeled with T = 10.0 --
50.0 eV and galactic N$_{\rm H}$ = 10$^{20}$ -- 10$^{21}$ cm$^{-2}$ in
the direction of the LMC. XBs in general show a power law spectrum
with N$_{\rm H}$ up to 10$^{22}$ cm$^{-2}$ because of intrinsic
absorption N$_{\rm H}$. So for most of the point and point like X-ray
sources PSPC count rates can be converted into HRI count rates by
dividing by a typical value of 3, though for very soft sources this
scale factor can be larger.    
Sources in regions with extended emission (e.g.\ 30 Dor or
N44) or close to another source can not always be resolved in PSPC
data and may result in false large converting factor. 

\begin{figure}
\centerline{\psfig{figure=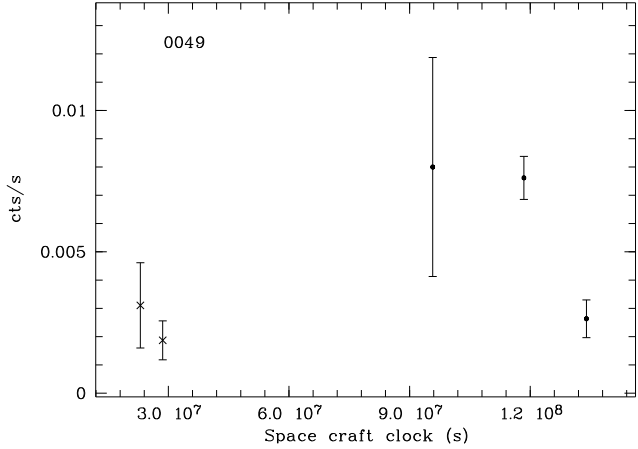,angle=270,width=9cm}}
\caption[]{\label{lc0049} Lightcurve of source No 49. 
Crosses for converted PSPC count rates, dots for HRI count rates. Zero
point of the space craft clock is 1990, June 21 21:06:50 UT.}  
\end{figure}

$\chi^{2}$ and the flux factor were again calculated for all
lightcurves including PSPC count rates (divided by 3.0) and upper
limits. Finally 26 sources show significant variability with reduced
$\chi^{2} >$ 5 corresponding to a probability $>$ 0.9999 (see Table
\ref{variable}).  
Four of them are new classified HRI sources (for sources No 49 and 364
see Sec.\,\ref{classhard}, for No 300 and 313 see Sec.\,\ref{stellar}).
As example the lightcurve of source No 49, a new HRI candidate for
a variable X-ray binary or AGN is shown in Fig.\,\ref{lc0049}.  

\begin{figure}
\centerline{\psfig{figure=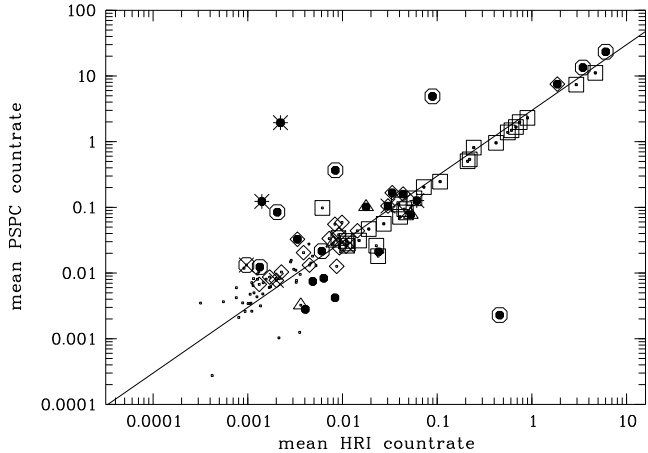,angle=270,width=9cm}}
\caption[]{\label{rate} Mean of observation-averaged count rates from
PSPC pointings over mean of observation-averaged count rates from HRI
pointings for ROSAT sources.  
Squares indicate SNRs, lozenges stars, hexagons XBs, triangles AGN,
and asterisks SSSs. Crossed symbols are already known candidates. 
Variable sources are additionally marked
with filled dots. The line indicates a PSPC/HRI conversion factor of 3.}
\end{figure}

PSPC count rates were determined in as many pointings as possible.
The mean value was calculated from these count rates and compared to
the HRI mean count rates (see Fig.\,\ref{rate}). The resulting
conversion factor is close to 3.0, only variable sources marked with
dots show bigger deviation.  

\section{Source classes}

In section \ref{identifiedsou} we discuss HRI sources which were
identified either with sources already known from literature or with
candidates which were found in former X-ray studies and in PSPC data
(HP99b). Section \ref{newclass} deals with new classification of HRI
sources based on their X-ray properties.

\subsection{Source identification}\label{identifiedsou}

For 97 HRI sources out of 138 which were also detected by the PSPC the
HRI observation yielded smaller positional error circles and
consequently more accurate source positions compared to the PSPC
results. Therefore for several sources likely optical counterparts
could be determined which was not possible only with PSPC data.  
 
94 HRI sources were identified with known objects in the LMC,
foreground stars, or background objects mainly
based on their position (see
Sec.\,\ref{correlate}). As they comprise different source types
X-ray properties characteristic for each source class could be derived from
HRI and PSPC data. Table \ref{identified} lists HRI sources
with identification. 

HP99b have shown that extent and extent likelihood as well as the
hardness ratios measured by the PSPC have characteristic values for
different source classes and can be used as classification criteria.

\begin{figure}
\centerline{\psfig{figure=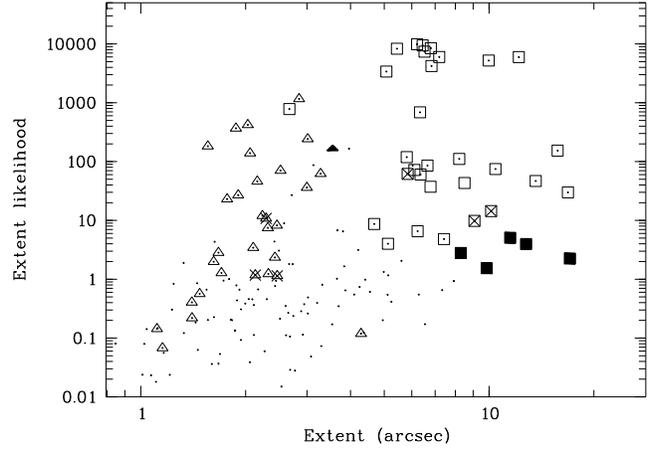,angle=270,width=9cm}}
\caption[]{\label{extent} Source extent and extent likelihood of HRI
sources in the LMC. SNRs are marked with open squares, known point
sources with open triangles. Crossed symbols are candidates for SNRs
or point sources known from literature, filled symbols are new
classifications.}
\end{figure}

In Fig.\,\ref{extent} extent and extent likelihood of the HRI sources
are shown. The extent was calculated in the maximum likelihood
algorithm and so gives the value resulting from fitting
Gaussians. Thus in some cases it may not be the extent of the whole
source but only of knots which were found within the extended source.
Identified SNRs, marked with open squares, are distributed in the region
with large extent and high extent likelihood. Crossed squares indicate
known SNR candidates and filled squares sources newly classified as
SNR candidates in this work. 
Point sources have lower extent likelihood unless they were extremely
bright like AB Dor (No 180), LMC X-1 (No 311), or RX J0439.8-6809 (No
4) where the deviation of the point spread function from the assumed 
Gaussian profile becomes significant. 

\subsubsection{Foreground stars}

By cross-correlating the HRI source catalogue with SIMBAD and
TYCHO catalogues and using the finding charts presented by Schmidtke
et al.\ (1994, hereafter SCF94), Cowley at al.\ (1997, hereafter
CSM97), and Schmidtke et al.\ (1999, hereafter SCC99)
39 sources were identified with Galactic foreground stars (Table
\ref{identified}). Most of them could also be identified with the
help of UBV photometry results presented by Gochermann \et\ (1993) and
Grothues \et\ (1997). On DSS-images there are point sources as very
likely optical counterparts at the positions of these HRI sources
within the error circle.   

Based on hardness ratios of the PSPC observations two point sources
were suggested as foreground star candidates by HP99b (No 189 and 349).
They were detected in PSPC images and their hardness ratios
are within the range characteristic of stars (HP99b). DSS images show
an optical point source within the improved HRI error circle in both
cases. 

\subsubsection{Supernova Remnants}

Most SNRs in the LMC are extended X-ray sources which could be resolved
by the HRI. They typically show extents of about 5\arcsec\ --
20\arcsec\ and high extent likelihood ($>$ 10.0). A total of 24 known
SNRs were observed by the HRI, four HRI sources are identified with
known SNR candidates (No 50, 231, 310, and 315). For both No
231 and 310 the measured hardness ratios are typical for SNRs. No 50
has a harder X-ray spectrum with HR1 = 1.00$\pm$0.10 and HR2 =
0.34$\pm$0.07. 

\subsubsection{Supersoft sources}

SSSs have very soft X-ray spectra and so far seven SSSs have been
discovered in the LMC (HP99b). Two of them were sources of the Einstein
LMC survey (Long et al.\ 1981) and five were found with the help of
the ROSAT PSPC. In the HRI pointings five LMC SSSs listed in Table
\ref{identified} were observed and detected with high existence
likelihood.  

\subsubsection{X-ray binaries}

Characteristic for most X-ray binaries is the hard X-ray spectrum and
flux variability. In HRI observations nine bright sources could be
identified with well known massive X-ray binaries (HMXB).
The point source RX J0532.7-6926, here No 238, has been suggested to
be a low mass X-ray binary (LMXB) candidate by Haberl \& Pietsch
(1999a, hereafter HP99a) and was also detected by the HRI. In HP99a a
lightcurve with PSPC and HRI measurements is presented and variability
is discussed in detail. Between 1990 and 1993 the source showed an
exponential intensity decay. 

\subsubsection{AGN}

Nine known background AGN with redshifts between 0.06 and 0.44 (SCF94;
CSM97; Crampton et al.\ 1997) were re-identified in the HRI pointings. 
Because of its positional coincidence with the radio source PKS 0552-640
and its hardness ratios measured by the PSPC the HRI source No 389 was
classified as AGN candidate (No 37 in HP99b). On the DSS frame an
optical source with m$_{B}$ = 16.3 within the HRI error circle is
identified as the most likely optical counterpart.

\subsection{New classifications}\label{newclass}

The extensive detection list produced from the HRI pointings towards
the LMC allowed us to search for new candidates for different source
types. In the course of studying the newly discovered HRI sources the
following parameters were of prime importance: count rates, source
extent, extent likelihood, flux variability, and counterparts in other
wavelengths. 

In addition to these X-ray properties we calculated the X-ray to optical
flux ratio of HRI sources, for which possible optical counterparts could
be found. The flux ratio was computed according to 
the equation log(f$_{x}$/f$_{opt}$) = log(3 $\cdot$ HRI counts/s
$\cdot 10^{-11}$) + 0.4 m$_{B}$ + 5.37 (Maccacaro et al.\ 1988; HP99b). 
The relation used for PSPC observations in HP99b was applied here for
HRI sources converting the HRI count rates to PSPC count rates by
multiplying by the factor of 3 which is typical for hard sources.
B magnitudes from the USNO-A1.0 Catalogue produced by the
United States Naval Observatory (Monet 1996) were used.
For several sources the optical counterpart could not be determined
uniquely. In such a case the magnitude of the brightest optical object
within the error circle was used resulting in lower
limits for log(f$_{x}$/f$_{opt}$). For the SNRs log(f$_{x}$/f$_{opt}$)
in general gives no quantitative information, but is an indicator that
this source class is bright in X-ray (log(f$_{x}$/f$_{opt}$) $>$ --1). 

\begin{figure}[t]
\centerline{\psfig{figure=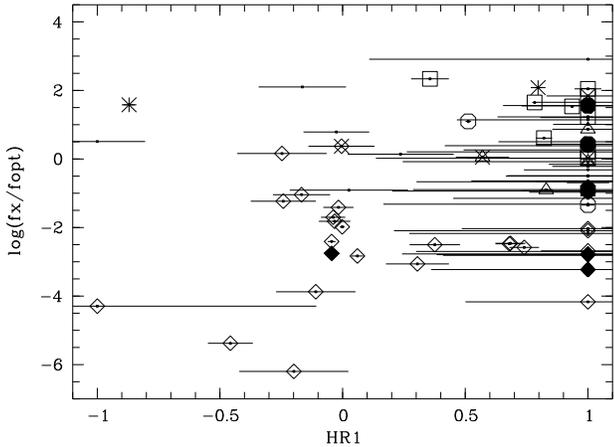,angle=270,width=9cm}}
\caption[]{\label{logfxfopt} Flux ratio log(f$_{x}$/f$_{opt}$) as a
function of hardness ratio 1. Open squares are SNRs, open
lozenges stars, open hexagons XBs, open triangles AGN, and asterisks
SSSs. Crossed symbols are already known
candidates and filled symbols are new classifications.} 
\end{figure}

As one can see in Fig.\,\ref{logfxfopt} stars in general have negative
log(f$_{x}$/f$_{opt}$), for AGN it is around zero, and for SSSs and XBs
it is mostly positive in particular when they were observed in their
X-ray active phase. Combination of f$_{x}$/f$_{opt}$ and the hardness
ratios provides a tool to exclude foreground stars. 

Newly discovered HRI sources which are suggested as candidates for
different source classes in this work can be found in Table
\ref{classified} and are discussed in the following. 

\subsubsection{SNR candidates}

Investigating the extent five HRI sources (No 197, 284, 288,
307, 338) not classified with the help of PSPC observations are suggested
as SNR candidates as their extent is larger than 8\arcsec\ (see
Fig.\,\ref{extent}).
Since they were not detected by the PSPC because of short exposure
times there is no spectral information about these sources which might
be crucial for further improvement of the classification. 

\subsubsection{Sources classified as stellar}\label{stellar}

For 11 HRI sources probable optical counterparts were found within the
error circle which are all bright (m$_{B} \leq$ 12.5), and their
log(f$_{x}$/f$_{opt}$) is negative ($<$ --2.0). For this reason these
sources are classified as stellar objects, and in particular the
brightest objects are likely foreground stars. Four sources were also
observed by the PSPC (No 90, 135, 217, 313), but as the errors of
their hardness ratios are large, no spectral information is given.

The lightcurve of No 300 shows a strong decrease of the X-ray
emission with a factor of 10 in 2 years indicating that the HRI
observations were performed after an emission maximum. 
The point source in the optical DSS image at the HRI position is very
likely the optical counterpart with a B magnitude
of m$_{B}$ = 12.4 according to the USNO-A1.0 Catalogue and
log(f$_{x}$/f$_{opt}$) = --2.75.

\subsubsection{LMC stars as candidates for high mass X-ray binaries}

Two X-ray point sources detected by the HRI were  
identified with known LMC O and B stars (No 328, Sanduleak 1970,
m$_{B}$ = 18.8 and No 332, Brunet \et\ 1975, m$_{B}$ = 13.6) because
of the positional coincidence. With HRI data no variability
investigations could be carried out for these X-ray sources, though
there exist many pointings in their direction, because they were both
detected only once and in other pointings the upper limits were
too high for this purpose. But their identification with optically
selected LMC stars allows us to classify them as candidates for high
mass X-ray binaries.  

\subsubsection{Sources with hard X-ray spectrum: Candidates for AGN or
X-ray binary} \label{classhard}

With the help of the hardness ratios and other characteristics
measured by the HRI like flux variability or f$_{x}$/f$_{opt}$
three HRI sources which were also detected by the PSPC 
could be classified as candidates either for X-ray binary or for AGN.

The point source No 49 shows significant flux variations, as it is shown in
Fig.\,\ref{lc0049}, and has a hard and/or highly absorbed X-ray spectrum
(HR1 = 1.00$\pm$0.71, HR2 = 0.26$\pm$0.16). On the DSS image a likely optical
counterpart with a B magnitude of 16.4 (according to the USNO-A1.0
Catalogue) is found. Therefore this source has been classified as
a candidate either for an X-ray binary or AGN.

Sources No 230 and 364 are further candidates for X-ray binary or
AGN as they have a hard and/or absorbed X-ray spectrum (HR1 =
1.00$\pm$0.35, HR2 = 1.00$\pm$0.98 and HR1 = 1.00$\pm$0.21, HR2 =
1.00$\pm$0.60 respectively).
Since source No 230 has a small positional error a probable
optical counterpart can be found on the DSS image. This counterpart is
faint (m$_{B}$ = 22.6), and we obtain a high log(f$_{x}$/f$_{opt}$) of
1.56.  
For source No 364 there is a relatively faint optical source
(m$_{B}$ = 18.2) within the error circle which might be the
counterpart (log(f$_{x}$/f$_{opt}$) = 0.43).

Another nine sources detected by the HRI were identified with sources in
the PSPC catalogue (HP99b) showing a hard X-ray spectrum. But from the
HRI observations no additional information could be obtained. Thus the
HRI sources are simply classified as hard X-ray sources because of the
hardness ratios of their PSPC detections. 

\subsection{Source distribution}

\begin{figure}
\begin{center}
\begin{tabular}{c}
\\[2cm]
\mbox{\psfig{figure=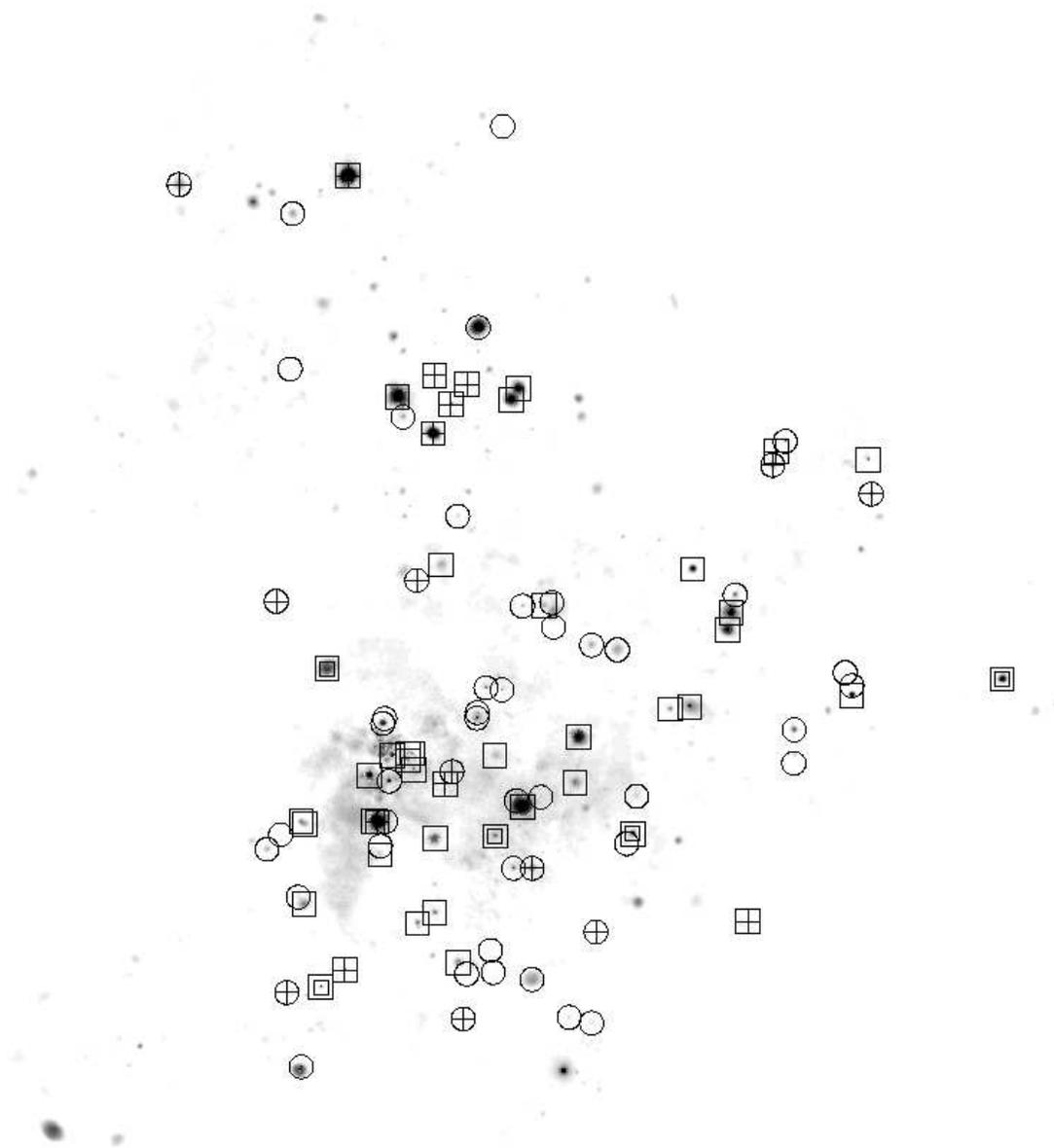,angle=0,width=175mm}}
\\[5mm]\\
\end{tabular}
\end{center}
\caption[]{\label{known} Identified HRI sources are plotted on a grey
scale PSPC image (0.1 -- 2.4 keV) of the LMC. Squares are SNRs,
circles are foreground stars, double squares SSSs, crossed squares
XBs, crossed circles AGN. Candidates from literature are included for
each source class.}
\end{figure}

\begin{figure}
\begin{center}
\begin{tabular}{c}
\\[2cm]
\mbox{\psfig{figure=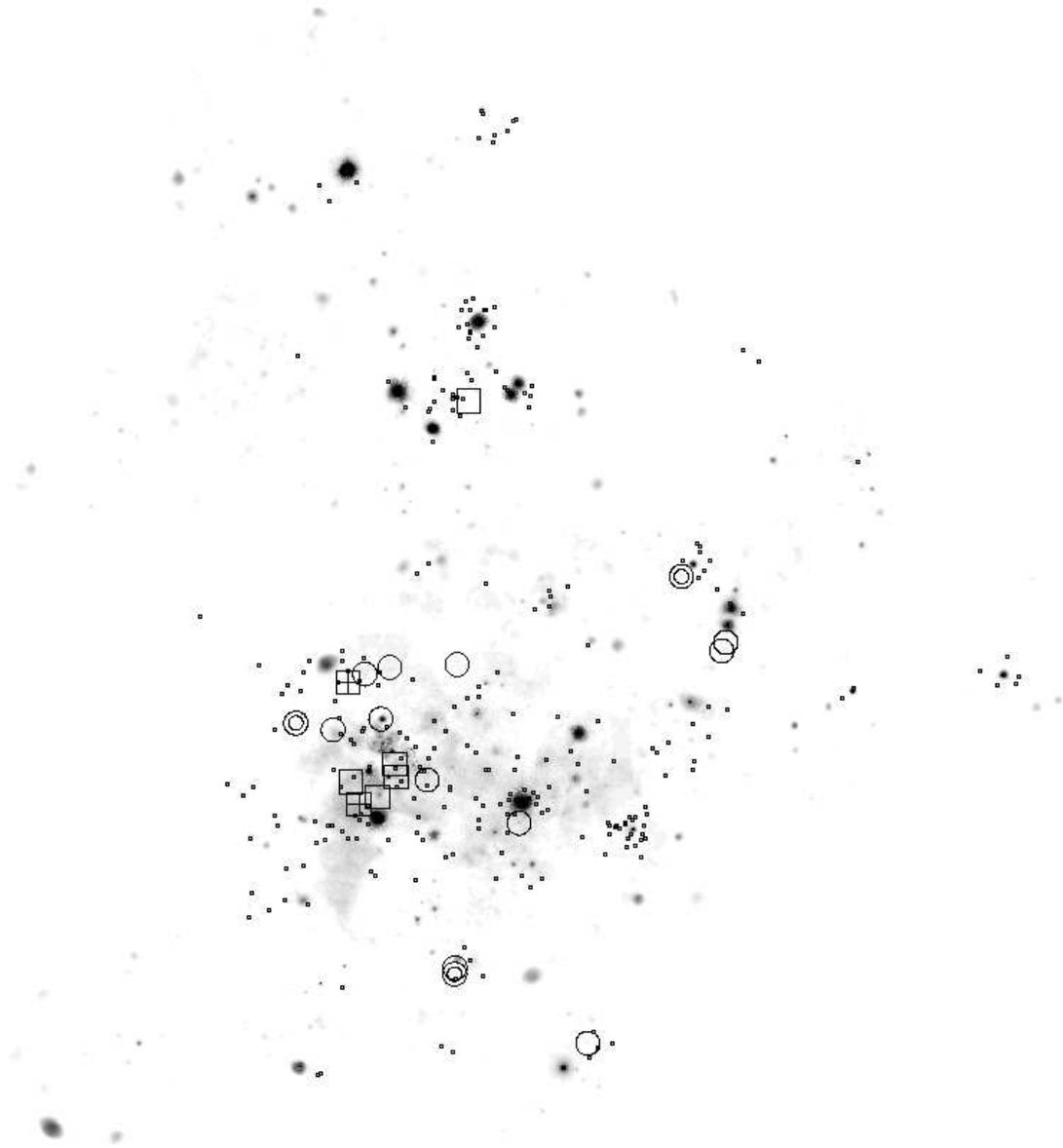,angle=0,width=175mm}}
\\[5mm]\\
\end{tabular}
\end{center}
\caption[]{\label{unidentified} The distribution of unidentified HRI
sources and new source classifications is shown. Unidentified HRI
sources are plotted as dots, squares are sources classified as SNR
candidates, circles as stellar sources, crossed squares as XB
candidates and double circles as candidates for XB or AGN.}
\end{figure}

Due to the high spatial resolution of the HRI many sources could be
detected both in the outer regions and in the optical bar region of
the LMC.
In Fig.\,\ref{known} HRI sources identified with known objects and
known candidates are plotted on a grey scale PSPC image (0.1 -- 2.4
keV) of the LMC (from HP99b). The sources are located in different
regions of the 
LMC and show no spatial preferences, it is not only background AGN or
foreground stars and candidates which are distributed over the whole
LMC region. There are still more than 250 non-identified point sources
which are homogeneously distributed in all LMC regions which were covered
by ROSAT HRI pointings as it is shown in Fig.\,\ref{unidentified}.  
In contrast, in PSPC observations not many additional sources could be
detected in the regions with strong diffuse emission, because the lower
spatial resolution hindered in distinguishing between extended and
point like emission (HP99b).

The HRI allows to study the extent of the sources to scales of
arcseconds. Therefore SNR candidates could be
found not only in regions without surrounding diffuse
emission. Four out of five newly suggested SNR candidates are located
in regions with diffuse emission between 30 Dor and LMC X-1 (see
Fig.\,\ref{unidentified}).  

Within and around the optical bar region several new stellar sources
and candidates for X-ray binary or AGN were found.

\section{Summary}

The analysis of all 543 ROSAT HRI pointed observations performed
between 1990 and 1998 with exposure times higher than 50 sec is
presented. Using a maximum likelihood algorithm the
source detection resulted in a catalogue of 397 sources which was
cross-correlated with the SIMBAD data base and the TYCHO catalogue.  
Further X-ray properties could be obtained for HRI sources contained
in the PSPC catalogue of HP99b. 

The high spatial resolution of the HRI enabled
the identification of 94 HRI sources with well known objects based on the
positional coincidence and considering their extent and hardness
ratios. The coordinates of most of the identified
sources could be improved to more accurate positions and allowed the
positional correction of other HRI sources. Thus for 254 sources the
systematic error for their coordinates could be reduced to values
smaller than 7\arcsec\ which is the standard systematic error of ROSAT
observations. 

For different source classes like SSS, X-ray binary, SNR,
Galactic stars, and background AGN classification
criteria could be derived from the extent and hardness ratios of the
identified sources. We looked for flux variability of the
sources and for likely optical counterparts. Five newly detected HRI
sources were classified as candidates for SNRs because of their
extent, two HRI sources which were identified with an LMC O and a B
star as HMXB candidates. Eleven sources with probable bright optical
counterpart and small X-ray to optical flux ratio are classified as
stellar sources. Three sources with hard and/or absorbed X-ray
spectrum indicated by the PSPC hardness ratios are likely candidates
for X-ray binaries or AGN. Two of the hard X-ray sources show flux
variability and for each of these an optical counterpart was found. 

With the help of HRI observations many new X-ray sources were found. 
Further follow-up observations in X-ray, optical, or radio
wavelengths with spectral information are needed to characterize these
sources in more detail.  

\acknowledgements
The ROSAT project is supported by the German
Bundesministerium f\"ur Bildung und Forschung (BMBF) and 
the Max-Planck-Gesellschaft. This research has been carried out by
making extensive use of the SIMBAD data base operated at CDS,
Strasbourg, France. 

\onecolumn

\clearpage

\clearpage

\clearpage

\begin{landscape}

\begin{table}
\scriptsize

\caption[]{Identified HRI sources in the LMC}
\begin{tabular}{rrrrrccrrccl}
\hline\noalign{\smallskip}
1~ & \multicolumn{1}{c}{2} & \multicolumn{1}{c}{3} & 4~ & \multicolumn{1}{c}{5} & 6 & 7 & \multicolumn{1}{c}{8} & \multicolumn{1}{c}{9} & 10 & 11 & ~~~12 \\
\hline\noalign{\smallskip}  
No & \multicolumn{1}{c}{RA} & \multicolumn{1}{c}{Dec} & \perr & \exil
& Count rate & \ext & \extl & \multicolumn{1}{c}{No} & HR1 & HR2 & Remarks \\
   & \multicolumn{2}{c}{(J2000.0)} & [\arcsec] & & [\ct] & [\arcsec] &
& \multicolumn{1}{c}{PSPC} & & & \\ 
\noalign{\smallskip}\hline\noalign{\smallskip}
  9 & 04 53 43.5 & -68 24 23 &  4.4 &    15.3 & 5.99e-04$\pm$1.63e-04 &  1.2$\pm$ 1.9 &   0.1 &     &  &  & foreground star G0 HD 268717 [GGO93]                                             \\ 
 12 & 04 54 30.2 & -68 18 01 &  9.1 &    12.7 & 1.30e-03$\pm$3.20e-04 &  0.0$\pm$ 0.0 &   0.0 &     &  &  & foreground star HD 31961                                                         \\ 
 15 & 04 58 25.1 & -69 08 22 &  4.1 &    39.8 & 5.77e-03$\pm$1.31e-03 &  2.3$\pm$ 2.0 &   1.2 &  816 &  1.00$\pm$0.51 &  0.29$\pm$0.13 & foreground star F7V [CSM97]                                                      \\ 
 16 & 04 58 43.9 & -68 50 50 &  4.1 &   317.5 & 4.48e-02$\pm$3.74e-03 &  0.0$\pm$ 0.0 &   0.0 &  742 & -0.04$\pm$0.05 &  0.09$\pm$0.07 & foreground star dMe HD 268840 [CSM97]                                            \\ 
 18 & 05 02 09.3 & -66 20 36 &  5.3 &    16.2 & 7.81e-03$\pm$3.22e-03 &  0.4$\pm$ 2.7 & 0.0&  304 &  0.68$\pm$0.06 &  0.08$\pm$0.08 & foreground star K0III [SCF94]                                                    \\
\hline
\end{tabular}

\vspace{2mm}
Notes to column No 9 (Table \ref{identified}), No 10 (Tables
\ref{classified} and \ref{wholecat}): 
Catalogue number from [HP99b].

\vspace{1mm}
Notes to column No 12 (Table \ref{identified}), No 13
(Tables \ref{classified} and \ref{wholecat}):

fg: foreground.

Candidates from literature are marked with ? behind the source class. 

New classification of this work are put in $<$ $>$.

Abbreviations for references in square brackets are given in the
literature list.
\end{table}

\clearpage

\addtocounter{table}{-1}
\begin{table}
\scriptsize
\caption[]{Continued}
\begin{tabular}{rrrrrccrrccl}
\hline\noalign{\smallskip}
1~ & \multicolumn{1}{c}{2} & \multicolumn{1}{c}{3} & 4~ & \multicolumn{1}{c}{5} & 6 & 7 & \multicolumn{1}{c}{8} & \multicolumn{1}{c}{9} & 10 & 11 & ~~~12 \\
\hline\noalign{\smallskip}  
No & \multicolumn{1}{c}{RA} & \multicolumn{1}{c}{Dec} & \perr & \exil & Count rate & \ext & \extl & \multicolumn{1}{c}{No} & HR1 & HR2 & Remarks \\
   & \multicolumn{2}{c}{(J2000.0)} & [\arcsec] & & [\ct] & [\arcsec] &
& \multicolumn{1}{c}{PSPC} & & & \\ 
\noalign{\smallskip}\hline\noalign{\smallskip}
  8 & 04 53 38.0 & -68 29 20 &  3.2 &   392.8 & 1.66e-02$\pm$7.28e-04 &  6.3$\pm$ 1.4 & 687.9 &  670 &  0.86$\pm$0.01 & -0.36$\pm$0.01 & SNR 0453-68.5                                                                    \\ 
 13 & 04 54 47.3 & -66 25 44 &  8.3 &    96.2 & 8.53e-03$\pm$7.03e-04 & 13.6$\pm$ 4.5 &  46.7 &  329 &  0.84$\pm$0.04 & -0.24$\pm$0.05 & SNR LHA 120-N 11L                                                                \\ 
 25 & 05 05 42.0 & -67 52 29 &  7.0 &  2223.3 & 5.83e-02$\pm$1.40e-03 &  6.8$\pm$ 1.1 &4214.3 &  592 &  0.81$\pm$0.01 & -0.21$\pm$0.01 & SNR DEM L 71                                                                     \\ 
 27 & 05 05 55.5 & -68 01 51 &  7.0 &  7040.7 & 2.16e-01$\pm$3.25e-03 & 12.2$\pm$ 1.7 &5968.8 &  614 &  0.89$\pm$0.01 & -0.19$\pm$0.02 & SNR LHA 120-N 23                                                                 \\ 
 43 & 05 08 58.6 & -68 43 35 &  7.0 & 13431.0 & 5.04e-01$\pm$7.44e-03 &  5.4$\pm$ 0.8 &8288.7 &  707 &  0.97$\pm$0.00 &  0.13$\pm$0.01 & SNR LHA 120-N 103B                                                               \\
\hline
\end{tabular}
\end{table}

\clearpage

\addtocounter{table}{-1}
\begin{table}
\scriptsize
\caption[]{Continued}
\begin{tabular}{rrrrrccrrccl}
\hline\noalign{\smallskip}
1~ & \multicolumn{1}{c}{2} & \multicolumn{1}{c}{3} & 4~ & \multicolumn{1}{c}{5} & 6 & 7 & \multicolumn{1}{c}{8} & \multicolumn{1}{c}{9} & 10 & 11 & ~~~12 \\
\hline\noalign{\smallskip}  
No & \multicolumn{1}{c}{RA} & \multicolumn{1}{c}{Dec} & \perr & \exil & Count rate & \ext & \extl & \multicolumn{1}{c}{No} & HR1 & HR2 & Remarks \\
   & \multicolumn{2}{c}{(J2000.0)} & [\arcsec] & & [\ct] & [\arcsec] &
& \multicolumn{1}{c}{PSPC} & & & \\ 
\noalign{\smallskip}\hline\noalign{\smallskip}
 10 & 04 54 10.7 & -66 43 17 &  7.3 &    55.1 & 8.01e-03$\pm$1.82e-03 &  1.5$\pm$ 1.5 &   0.6 &  411 &  1.00$\pm$0.24 &  0.29$\pm$0.07 & AGN RX J0454.2-6643, z=0.228 [CGC97]                                            \\ 
 20 & 05 03 04.0 & -66 33 44 &  2.2 &   188.8 & 5.35e-02$\pm$8.03e-03 &  0.0$\pm$ 0.0 &   0.0 &  380 &  0.83$\pm$0.03 &  0.17$\pm$0.03 & AGN RX J0503.1-6634, z=0.064 [SCF94]                                             \\ 
 86 & 05 17 16.9 & -70 44 01 &  2.2 &   186.4 & 1.03e-02$\pm$1.24e-03 &  2.5$\pm$ 1.4 &   8.2 &     &  &  & AGN RX J0517.3-7044, z=0.169 [CSM97]                                     \\ 
124 & 05 24 02.5 & -70 11 09 &  1.6 &   800.1 & 1.57e-02$\pm$8.88e-04 &  1.6$\pm$ 1.2 &   2.0 & 1094 &  0.91$\pm$0.02 &  0.27$\pm$0.04 & AGN RX J0524.0-7011, z=0.151 [SCF94]                                             \\ 
220 & 05 31 31.8 & -71 29 46 &  3.5 &    46.4 & 1.19e-02$\pm$3.54e-03 &  0.0$\pm$ 0.0 &   0.0 &     &  &  & AGN RX J0531.5-7130, z=0.221 [SCF94]                                   \\
\hline
\end{tabular}\label{identified}

\caption[]{Classified HRI sources}
\begin{tabular}{rrrrrccrrrccl}
\hline\noalign{\smallskip}
1~ & \multicolumn{1}{c}{2} & \multicolumn{1}{c}{3} & 4~ &
\multicolumn{1}{c}{5} & 6 & 7 & \multicolumn{1}{c}{8} &
\multicolumn{1}{c}{9} & \multicolumn{1}{c}{10} & 11 & 12 & ~~~13 \\
\hline\noalign{\smallskip}  
No & \multicolumn{1}{c}{RA} & \multicolumn{1}{c}{Dec} & \perr & \exil
& Count rate & \ext & \extl & log(f$_{x}$/f$_{opt}$) & \multicolumn{1}{c}{No} & HR1 & HR2 & Remarks \\
   & \multicolumn{2}{c}{(J2000.0)} & [\arcsec] & & [\ct] & [\arcsec] &
& & \multicolumn{1}{c}{PSPC} & & & \\ 
\noalign{\smallskip}\hline\noalign{\smallskip}
 28 & 05 05 58.4 & -68 10 30 &  8.4 &    22.8 & 1.85e-03$\pm$4.21e-04 &  2.7$\pm$ 3.4 &   0.2 &-3.09 &     &      &      & $<$stellar$>$                                                                   \\
 29 & 05 06 16.7 & -68 15 09 &  9.7 &    30.5 & 3.82e-03$\pm$6.81e-04 &  0.0$\pm$ 0.0 &   0.0 &-2.77 & &  & & $<$stellar$>$                                                     \\  
 49 & 05 10 28.7 & -67 37 41 &  7.1 &   484.0 & 5.01e-03$\pm$3.10e-04 &  0.0$\pm$ 0.0 &   0.0 &-0.89 &  559 &  1.00$\pm$0.71 &  0.26$\pm$0.16 & $<$XB$>$ or $<$AGN$>$ \\  
 90 & 05 17 47.8 & -71 44 05 & 11.4 &    15.9 & 1.96e-03$\pm$6.46e-04 &  1.9$\pm$ 2.3 &   0.3 &-3.06 & 1305 &  1.00$\pm$1.34 &  0.45$\pm$0.15 & $<$stellar$>$                            \\  
135 & 05 25 22.5 & -69 49 16 &  4.9 &    29.6 & 2.12e-03$\pm$3.86e-04 &  0.0$\pm$ 0.0 &   0.0 &-3.23 & 1025 &  1.00$\pm$0.64 &  1.00$\pm$2.40 & $<$stellar$>$                         \\
\hline
\end{tabular}
\label{classified} 

\end{table}

\clearpage

\begin{table}
\scriptsize
\caption[]{\label{wholecat}HRI sources in the LMC region}
\begin{tabular}{rrrrrccrrrccl}
\hline\noalign{\smallskip}
1~ & \multicolumn{1}{c}{2} & \multicolumn{1}{c}{3} & 4~ &
\multicolumn{1}{c}{5} & 6 & 7 & \multicolumn{1}{c}{8} &
\multicolumn{1}{c}{9} & \multicolumn{1}{c}{10} & 11 & 12 & ~~~13 \\
\hline\noalign{\smallskip}  
No & \multicolumn{1}{c}{RA} & \multicolumn{1}{c}{Dec} & \perr & \exil
& Count rate & \ext & \extl & Count rate & \multicolumn{1}{c}{No} & HR1 & HR2 & Remarks \\
 & & & & & & & & \multicolumn{1}{c}{PSPC} & \multicolumn{1}{c}{PSPC} & & & \\ 
 & \multicolumn{2}{c}{(J2000.0)} & [\arcsec] & & [\ct] & [\arcsec] &
& \multicolumn{1}{c}{[\ct]} & & & & \\ 
\noalign{\smallskip}\hline\noalign{\smallskip}
  1 & 04 38 23.9 & -68 08 24 & 5.7 & 10.7 & 4.42e-03$\pm$1.82e-03 & 0.0$\pm$ 0.0 & 0.0 & $<$1.74E-03 & & & & \\ 
  2 & 04 38 31.3 & -68 12 01 & 3.0 & 32.7 & 2.81e-03$\pm$6.18e-04 & 0.0$\pm$ 0.0 & 0.0 & $<$1.71E-03 & & & & \\ 
  3 & 04 39 44.7 & -67 58 39 & 10.0 & 10.1 & 3.81e-03$\pm$1.21e-03 & 5.2$\pm$ 5.7 & 0.4 &  & & & & \\ 
  4 & 04 39 49.6 & -68 09 01 & 0.4 & 6228.5 & 1.66e-01$\pm$4.89e-03 & 2.1$\pm$ 0.6 & 138.9 & 1.15E+00  & 628 & -1.00$\pm$0.01 & & SSS RX J0439.8-6809  \\
\hline
\end{tabular}

\vspace{2mm}
Notes to column No 6: 
HRI count rate ist the output value from maximum likelihood algorithm
with the smallest positional error.

\vspace{1mm}
Notes to column No 9:
For sources which are listed in [HP99b] PSPC count rates are taken from
the PSPC catalogue. 
Otherwise PSPC count rate is the 95.4\% (2$\sigma$) upper limit from
maximum likelihood algorithm of the pointing with maximum exposure. 

\end{table}

\clearpage

\addtocounter{table}{-1}
\begin{table}
\scriptsize
\caption[]{Continued}
\begin{tabular}{rrrrrccrrrccl}
\hline\noalign{\smallskip}
1~ & \multicolumn{1}{c}{2} & \multicolumn{1}{c}{3} & 4~ &
\multicolumn{1}{c}{5} & 6 & 7 & \multicolumn{1}{c}{8} &
\multicolumn{1}{c}{9} & \multicolumn{1}{c}{10} & 11 & 12 & ~~~13 \\
\hline\noalign{\smallskip}  
No & \multicolumn{1}{c}{RA} & \multicolumn{1}{c}{Dec} & \perr & \exil
& Count rate & \ext & \extl & Count rate & \multicolumn{1}{c}{No} & HR1 & HR2 & Remarks \\
 & & & & & & & & \multicolumn{1}{c}{PSPC} & \multicolumn{1}{c}{PSPC} & & & \\ 
 & \multicolumn{2}{c}{(J2000.0)} & [\arcsec] & & [\ct] & [\arcsec] &
& \multicolumn{1}{c}{[\ct]} & & & & \\ 
\noalign{\smallskip}\hline\noalign{\smallskip}
 51 & 05 10 50.4 & -69 05 13 & 9.3 & 10.0 & 2.82e-03$\pm$1.08e-03 & 3.0$\pm$ 3.2 & 0.3 & $<$6.39E-03 & & & & \\ 
 52 & 05 10 56.1 & -69 22 14 & 7.9 & 16.5 & 3.49e-03$\pm$1.31e-03 & 0.0$\pm$ 0.0 & 0.0 & $<$8.88E-03 & & & & \\ 
 53 & 05 11 51.1 & -69 10 16 & 7.3 & 25.4 & 2.72e-03$\pm$5.46e-04 & 2.3$\pm$ 3.8 & 0.1 & $<$8.57E-03 & & & & \\ 
 54 & 05 12 21.3 & -69 08 29 & 7.5 & 10.7 & 1.11e-03$\pm$3.46e-04 & 0.0$\pm$ 0.0 & 0.0 & $<$3.21E-03 & & & & \\
\hline
\end{tabular}
\end{table}

\clearpage

\addtocounter{table}{-1}
\begin{table}
\scriptsize
\caption[]{Continued}
\begin{tabular}{rrrrrccrrrccl}
\hline\noalign{\smallskip}
1~ & \multicolumn{1}{c}{2} & \multicolumn{1}{c}{3} & 4~ &
\multicolumn{1}{c}{5} & 6 & 7 & \multicolumn{1}{c}{8} &
\multicolumn{1}{c}{9} & \multicolumn{1}{c}{10} & 11 & 12 & ~~~13 \\
\hline\noalign{\smallskip}  
No & \multicolumn{1}{c}{RA} & \multicolumn{1}{c}{Dec} & \perr & \exil
& Count rate & \ext & \extl & Count rate & \multicolumn{1}{c}{No} & HR1 & HR2 & Remarks \\
 & & & & & & & & \multicolumn{1}{c}{PSPC} & \multicolumn{1}{c}{PSPC} & & & \\ 
 & \multicolumn{2}{c}{(J2000.0)} & [\arcsec] & & [\ct] & [\arcsec] &
& \multicolumn{1}{c}{[\ct]} & & & & \\ 
\noalign{\smallskip}\hline\noalign{\smallskip}
101 & 05 20 54.9 & -67 44 59 & 7.0 & 14.7 & 2.95e-03$\pm$6.37e-04 & 0.0$\pm$ 0.0 & 0.0 & $<$2.21E-03 & & & & \\ 
102 & 05 21 39.2 & -68 53 43 & 13.8 & 10.9 & 2.86e-03$\pm$7.89e-04 & 2.9$\pm$ 7.7 & 0.0 & $<$3.85E-03 & & & & \\ 
103 & 05 22 08.2 & -68 04 28 & 2.4 & 124.0 & 5.61e-03$\pm$5.50e-04 & 0.0$\pm$ 0.0 & 0.0 & $<$4.40E-03 & & & & foreground star HD 35862 \\ 
104 & 05 22 19.6 & -67 51 31 & 2.7 & 30.5 & 1.72e-03$\pm$3.14e-04 & 1.4$\pm$ 1.5 & 0.4 & $<$1.69E-03 & & & & foreground star HD 269422  \\
\hline
\end{tabular}
\end{table}

\clearpage

\addtocounter{table}{-1}
\begin{table}
\scriptsize
\caption[]{Continued}
\begin{tabular}{rrrrrccrrrccl}
\hline\noalign{\smallskip}
1~ & \multicolumn{1}{c}{2} & \multicolumn{1}{c}{3} & 4~ &
\multicolumn{1}{c}{5} & 6 & 7 & \multicolumn{1}{c}{8} &
\multicolumn{1}{c}{9} & \multicolumn{1}{c}{10} & 11 & 12 & ~~~13 \\
\hline\noalign{\smallskip}  
No & \multicolumn{1}{c}{RA} & \multicolumn{1}{c}{Dec} & \perr & \exil
& Count rate & \ext & \extl & Count rate & \multicolumn{1}{c}{No} & HR1 & HR2 & Remarks \\
 & & & & & & & & \multicolumn{1}{c}{PSPC} & \multicolumn{1}{c}{PSPC} & & & \\ 
 & \multicolumn{2}{c}{(J2000.0)} & [\arcsec] & & [\ct] & [\arcsec] &
& \multicolumn{1}{c}{[\ct]} & & & & \\ 
\noalign{\smallskip}\hline\noalign{\smallskip}
151 & 05 26 31.5 & -66 01 04 & 8.1 & 11.1 & 5.62e-04$\pm$1.79e-04 & 0.0$\pm$ 0.0 & 0.0 & $<$2.46E-03 & & & & \\ 
152 & 05 26 32.1 & -69 44 48 & 8.6 & 10.5 & 2.12e-03$\pm$7.00e-04 & 0.0$\pm$ 0.0 & 0.0 & $<$1.28E-03 & & & & \\ 
153 & 05 26 32.9 & -69 54 12 & 4.7 & 18.1 & 6.00e-04$\pm$1.46e-04 & 0.0$\pm$ 0.0 & 0.0 & 2.46E-03  &1041 & 1.00$\pm$0.94 & 1.00$\pm$0.99 & \\ 
154 & 05 26 34.9 & -63 41 34 & 3.5 & 21.5 & 8.72e-04$\pm$2.44e-04 & 0.0$\pm$ 0.0 & 0.0 & $<$1.70E-02 & & & & foreground star HD 36355 \\
\hline
\end{tabular}
\end{table}

\clearpage

\addtocounter{table}{-1}
\begin{table}
\scriptsize
\caption[]{Continued}
\begin{tabular}{rrrrrccrrrccl}
\hline\noalign{\smallskip}
1~ & \multicolumn{1}{c}{2} & \multicolumn{1}{c}{3} & 4~ &
\multicolumn{1}{c}{5} & 6 & 7 & \multicolumn{1}{c}{8} &
\multicolumn{1}{c}{9} & \multicolumn{1}{c}{10} & 11 & 12 & ~~~13 \\
\hline\noalign{\smallskip}  
No & \multicolumn{1}{c}{RA} & \multicolumn{1}{c}{Dec} & \perr & \exil
& Count rate & \ext & \extl & Count rate & \multicolumn{1}{c}{No} & HR1 & HR2 & Remarks \\
 & & & & & & & & \multicolumn{1}{c}{PSPC} & \multicolumn{1}{c}{PSPC} & & & \\ 
 & \multicolumn{2}{c}{(J2000.0)} & [\arcsec] & & [\ct] & [\arcsec] &
& \multicolumn{1}{c}{[\ct]} & & & & \\ 
\noalign{\smallskip}\hline\noalign{\smallskip}
201 & 05 29 45.0 & -65 16 03 & 6.6 & 18.9 & 2.47e-03$\pm$4.65e-04 & 0.0$\pm$ 0.0 & 0.0 & $<$2.15E-03 & & & & \\ 
202 & 05 29 48.3 & -65 56 46 & 2.0 & 119.2 & 1.94e-03$\pm$2.05e-04 & 0.0$\pm$ 0.0 & 0.0 & 2.43E-01  & 204 & 0.84$\pm$0.04 & 0.49$\pm$0.06 & HMXB RXJ0529.8-6556 [HP99a] \\ 
203 & 05 30 04.7 & -65 20 08 & 5.4 & 17.3 & 1.81e-03$\pm$3.65e-04 & 0.0$\pm$ 0.0 & 0.0 & $<$3.64E-03 & & & & \\ 
204 & 05 30 09.3 & -66 06 56 & 6.7 & 10.2 & 6.24e-04$\pm$1.68e-04 & 0.0$\pm$ 0.0 & 0.0 & 2.19E-03  & 251 & 1.00$\pm$0.83 & 0.32$\pm$0.16 & \\
\hline
\end{tabular}
\end{table}

\clearpage

\addtocounter{table}{-1}
\begin{table}
\scriptsize
\caption[]{Continued}
\begin{tabular}{rrrrrccrrrccl}
\hline\noalign{\smallskip}
1~ & \multicolumn{1}{c}{2} & \multicolumn{1}{c}{3} & 4~ &
\multicolumn{1}{c}{5} & 6 & 7 & \multicolumn{1}{c}{8} &
\multicolumn{1}{c}{9} & \multicolumn{1}{c}{10} & 11 & 12 & ~~~13 \\
\hline\noalign{\smallskip}  
No & \multicolumn{1}{c}{RA} & \multicolumn{1}{c}{Dec} & \perr & \exil
& Count rate & \ext & \extl & Count rate & \multicolumn{1}{c}{No} & HR1 & HR2 & Remarks \\
 & & & & & & & & \multicolumn{1}{c}{PSPC} & \multicolumn{1}{c}{PSPC} & & & \\ 
 & \multicolumn{2}{c}{(J2000.0)} & [\arcsec] & & [\ct] & [\arcsec] &
& \multicolumn{1}{c}{[\ct]} & & & & \\ 
\noalign{\smallskip}\hline\noalign{\smallskip}
251 & 05 34 15.8 & -69 14 17 & 2.9 & 29.4 & 1.23e-03$\pm$2.52e-04 & 0.8$\pm$ 2.1 & 0.0 & $<$3.34E-03 & & & & \\ 
252 & 05 34 16.3 & -70 33 43 & 10.6 & 10.5 & 3.99e-03$\pm$1.08e-03 & 7.4$\pm$ 4.5 & 4.8 & 6.49E-02  &1160 & 0.82$\pm$0.03 & 0.08$\pm$0.04 & SNR DEM L 238  \\ 
253 & 05 34 26.4 & -69 28 56 & 6.1 & 10.4 & 2.25e-03$\pm$8.76e-04 & 0.0$\pm$ 0.0 & 0.0 & $<$1.46E-03 & & & & \\ 
254 & 05 34 27.7 & -69 25 40 & 5.9 & 11.0 & 2.65e-03$\pm$9.41e-04 & 0.0$\pm$ 0.0 & 0.0 & $<$8.05E-04 & & & & $<$stellar$>$  \\
\hline
\end{tabular}
\end{table}

\clearpage

\addtocounter{table}{-1}
\begin{table}
\scriptsize
\caption[]{Continued}
\begin{tabular}{rrrrrccrrrccl}
\hline\noalign{\smallskip}
1~ & \multicolumn{1}{c}{2} & \multicolumn{1}{c}{3} & 4~ &
\multicolumn{1}{c}{5} & 6 & 7 & \multicolumn{1}{c}{8} &
\multicolumn{1}{c}{9} & \multicolumn{1}{c}{10} & 11 & 12 & ~~~13 \\
\hline\noalign{\smallskip}  
No & \multicolumn{1}{c}{RA} & \multicolumn{1}{c}{Dec} & \perr & \exil
& Count rate & \ext & \extl & Count rate & \multicolumn{1}{c}{No} & HR1 & HR2 & Remarks \\
 & & & & & & & & \multicolumn{1}{c}{PSPC} & \multicolumn{1}{c}{PSPC} & & & \\ 
 & \multicolumn{2}{c}{(J2000.0)} & [\arcsec] & & [\ct] & [\arcsec] &
& \multicolumn{1}{c}{[\ct]} & & & & \\ 
\noalign{\smallskip}\hline\noalign{\smallskip}
301 & 05 38 43.7 & -69 06 05 & 0.7 & 1523.9 & 8.47e-03$\pm$3.04e-04 & 4.0$\pm$ 1.1 & 165.4 & & & & & cluster of stars HD 38268; RMC 136 \\ 
302 & 05 38 43.8 & -69 10 12 & 2.1 & 41.7 & 2.44e-03$\pm$1.98e-04 & 7.1$\pm$ 2.0 & 107.4 & & & & & knot in 30 Dor \\ 
303 & 05 38 46.7 & -69 02 25 & 2.9 & 33.4 & 2.25e-03$\pm$2.06e-04 & 8.1$\pm$ 2.6 & 71.9 & & & & & knot in 30 Dor \\ 
304 & 05 38 49.8 & -68 35 04 & 7.4 & 39.1 & 1.50e-03$\pm$2.80e-04 & 1.6$\pm$ 2.0 & 0.2 & & & & & \\
\hline
\end{tabular}
\end{table}

\clearpage

\addtocounter{table}{-1}
\begin{table}
\scriptsize
\caption[]{Continued}
\begin{tabular}{rrrrrccrrrccl}
\hline\noalign{\smallskip}
1~ & \multicolumn{1}{c}{2} & \multicolumn{1}{c}{3} & 4~ &
\multicolumn{1}{c}{5} & 6 & 7 & \multicolumn{1}{c}{8} &
\multicolumn{1}{c}{9} & \multicolumn{1}{c}{10} & 11 & 12 & ~~~13 \\
\hline\noalign{\smallskip}  
No & \multicolumn{1}{c}{RA} & \multicolumn{1}{c}{Dec} & \perr & \exil
& Count rate & \ext & \extl & Count rate & \multicolumn{1}{c}{No} & HR1 & HR2 & Remarks \\
 & & & & & & & & \multicolumn{1}{c}{PSPC} & \multicolumn{1}{c}{PSPC} & & & \\ 
 & \multicolumn{2}{c}{(J2000.0)} & [\arcsec] & & [\ct] & [\arcsec] &
& \multicolumn{1}{c}{[\ct]} & & & & \\ 
\noalign{\smallskip}\hline\noalign{\smallskip}
351 & 05 43 58.7 & -65 39 55 & 4.4 & 16.4 & 1.88e-03$\pm$6.11e-04 & 0.0$\pm$ 0.0 & 0.0 & 6.86E-03  & 148 & 1.00$\pm$0.87 & 1.00$\pm$0.66 & \\ 
352 & 05 44 06.0 & -71 00 51 & 7.8 & 12.3 & 1.34e-03$\pm$5.17e-04 & 0.0$\pm$ 0.0 & 0.0 & 4.41E-02  &1225 & 1.00$\pm$0.03 & 0.65$\pm$0.03 & HMXB RX J0544.1-7100 [HP99b] \\ 
353 & 05 44 10.5 & -69 46 49 & 7.9 & 11.3 & 6.22e-04$\pm$2.01e-04 & 0.0$\pm$ 0.0 & 0.0 & $<$2.21E-02 & & & & \\ 
354 & 05 44 29.3 & -71 11 55 & 10.5 & 11.8 & 1.73e-03$\pm$5.03e-04 & 0.0$\pm$ 0.0 & 0.0 & $<$1.51E-03 & & & & \\
\hline
\end{tabular}
\end{table}
\label{wholecat}

\end{landscape}

\end{document}